\input{aipcheck}

\documentclass[
    ,final            
  ]
  {aipproc}

\layoutstyle{6x9}

\begin{document}

\title{
Precision Measurement of Parity Violation in Polarized Cold Neutron Capture on the Proton:\newline
the NPD$\gamma$ Experiment}




\author{Bernhard Lauss}{
address={University of California, Berkeley and Lawrence Berkeley Nat. Lab., Berkeley, CA 94720, USA
}
}
\author{J.D.~Bowman}{
address={Los Alamos National Laboratory, Los Alamos, NM 87545, USA
}
}
\author{      R.~D.~Carlini}{
address={Thomas Jefferson National Accelerator Facility,  Newport News, VA 23606, USA
}
}
\author{      T.~E.~Chupp}{
address={Department of Physics, University of Michigan, Ann Arbor, MI 48109, USA
}
}
\author{      W.~Chen}{
address={Department of Physics, Indiana University, Bloomington, IN 47408, USA
}
}
\author{}{
address={National Institute of Standards and Technology, Gaithersburg, MD 20899, USA
}
}
\author{      S.~Corvig}{
address={Department of Physics, University of New Hampshire, Durham, NH 03824, USA
}
}
\author{      M.~Dabaghyan}{
address={Department of Physics, University of New Hampshire, Durham, NH 03824, USA
}
}
\author{      D.~Desai}{
address={Department of Physics, University of Tennessee, Knoxville, TN 37996, USA
}
}
\author{      S.~J.~Freedman}{
address={University of California, Berkeley and Lawrence Berkeley Nat. Lab., Berkeley, CA 94720, USA
}
}
\author{      T.~R.~Gentile}{
address={National Institute of Standards and Technology, Gaithersburg, MD 20899, USA
}
}
\author{      M.~T.~Gericke}{
address={Department of Physics, University of Manitoba, Winnipeg, Manitoba, R3T 2N2 Canada
}
}
\author{      R.~C.~Gillis}{
address={Department of Physics, University of Manitoba, Winnipeg, Manitoba, R3T 2N2 Canada
}
}
\author{      G.~L.~Greene}{
address={Department of Physics, University of Tennessee, Knoxville, TN 37996, USA
}
}
\author{}{
address={Oak Ridge National Laboratory, Oak Ridge, TN 37831, USA
}
}
\author{      F.~W.~Hersman}{
address={Department of Physics, University of New Hampshire, Durham, NH 03824, USA
}
}
\author{      T.~Ino}{
address={High Energy Accelerator Research Organization (KEK), Tsukuba-shi, 305-0801, Japan 
}
}
\author{      T.~Ito}{
address={Los Alamos National Laboratory, Los Alamos, NM 87545, USA
}
}
\author{      G.~L.~Jones}{
address={Department of Physics, Hamilton College, Clinton, NY 13323, USA
}
}
\author{      M.~Kandes}{
address={Department of Physics, University of Michigan, Ann Arbor, MI 48109, USA
}
}
\author{      M.~Leuschner}{
address={Department of Physics, Indiana University, Bloomington, IN 47408, USA 
}
}
\author{     B.~Lozowski}{
address={Indiana University Cyclotron Facility, Bloomington, IN 47408, USA
}
}
\author{      R.~Mahurin}{
address={Department of Physics, University of Tennessee, Knoxville, TN 37996, USA
}
}
\author{      M.~Mason}{
address={Department of Physics, University of New Hampshire, Durham, NH 03824, USA
}
}
\author{      Y.~Masuda}{
address={High Energy Accelerator Research Organization (KEK), Tsukuba-shi, 305-0801, Japan 
}
}
\author{      J.~Mei}{
address={Department of Physics, Indiana University, Bloomington, IN 47408, USA 
}
}
\author{      G.~S.~Mitchell}{
address={Los Alamos National Laboratory, Los Alamos, NM 87545, USA
}
}
\author{      S.~Muto}{
address={High Energy Accelerator Research Organization (KEK), Tsukuba-shi, 305-0801, Japan 
}
}
\author{      H.~Nann}{
address={Department of Physics, Indiana University, Bloomington, IN 47408, USA 
}
}
\author{      S.A.~Page}{
address={Department of Physics, University of Manitoba, Winnipeg, Manitoba, R3T 2N2 Canada
}
}
\author{      S.I.~Penttila}{
address={Los Alamos National Laboratory, Los Alamos, NM 87545, USA
}
}
\author{      W.D.~Ramsay}{
address={Department of Physics, University of Manitoba, Winnipeg, Manitoba, R3T 2N2 Canada
}
}
\author{}{
address={TRIUMF, Vancouver, British Columbia, V6T2A3 Canada
}
}
\author{      S.~Santra}{
address={Bhabha Atomic Research Center, Mumbai, India 
}
}
\author{      P-N.~Seo}{
address={Department of Physics, North Carolina State University, Raleigh, NC 27695, USA\newline
$^{\alpha}$Joint Institute of Nuclear Research, Dubna, Russia\newline
$^{\beta}$Department of Physics, University of Dayton, Dayton, OH 45469, USA
}
}
\author{      E.I.~Sharapov$^{\alpha}$}{
address={Department of Physics, North Carolina State University, Raleigh, NC 27695, USA\newline
$^{\alpha}$Joint Institute of Nuclear Research, Dubna, Russia\newline
$^{\beta}$Department of Physics, University of Dayton, Dayton, OH 45469, USA
}
}

\author{      T.B.~Smith$^{\beta}$}{
address={Department of Physics, North Carolina State University, Raleigh, NC 27695, USA\newline
$^{\alpha}$Joint Institute of Nuclear Research, Dubna, Russia\newline
$^{\beta}$Department of Physics, University of Dayton, Dayton, OH 45469, USA
}
}
\author{      W.M.~Snow}{
address={Department of Physics, Indiana University, Bloomington, IN 47408, USA 
}
}
\author{      W.S.~Wilburn}{
address={Los Alamos National Laboratory, Los Alamos, NM 87545, USA
}
}
\author{      V.~Yuan}{
address={Los Alamos National Laboratory, Los Alamos, NM 87545, USA
}
}
\author{     H.~Zhu}{
address={Department of Physics, University of New Hampshire, Durham, NH 03824, USA
}
}

\begin{abstract}
The NPD$\gamma$ experiment\footnote{\tt http://p23.lanl.gov/len/npdg/}
at the Los Alamos Neutron Science Center (LANSCE)
is dedicated to measure with high precision 
the parity violating asymmetry in the $\gamma$
emission after capture of spin polarized cold neutrons in para-hydrogen.
The measurement will determine unambiguously 
the weak pion-nucleon-nucleon ($\pi NN$) coupling constant {\it f$^1_{\pi}$}.
\end{abstract}

\maketitle

\begin{figure}[h]
{\includegraphics[height= 0.24\textheight ]{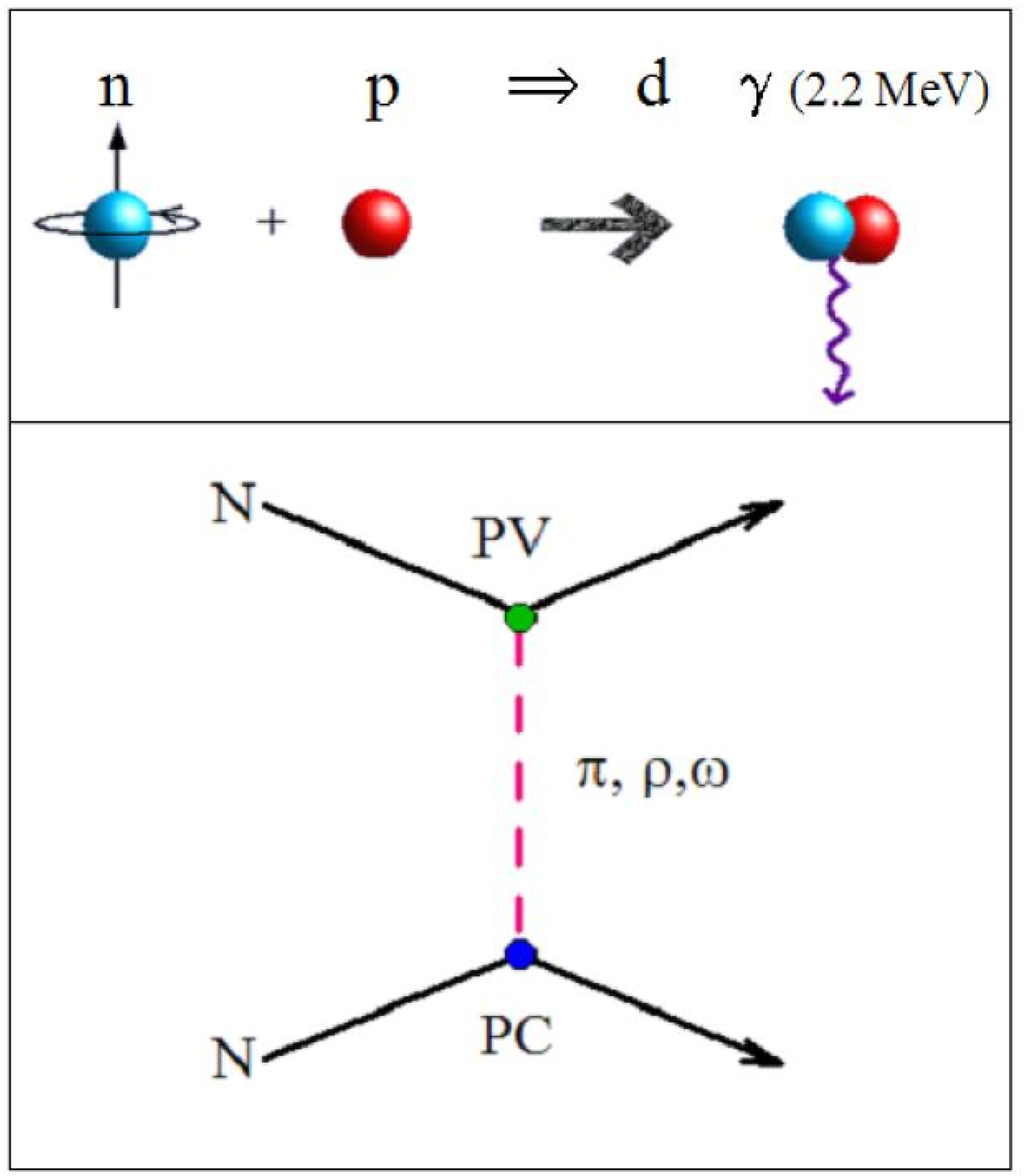}}
{\hspace*{10mm} \includegraphics[width=0.3\textwidth, height= 0.24\textheight ]{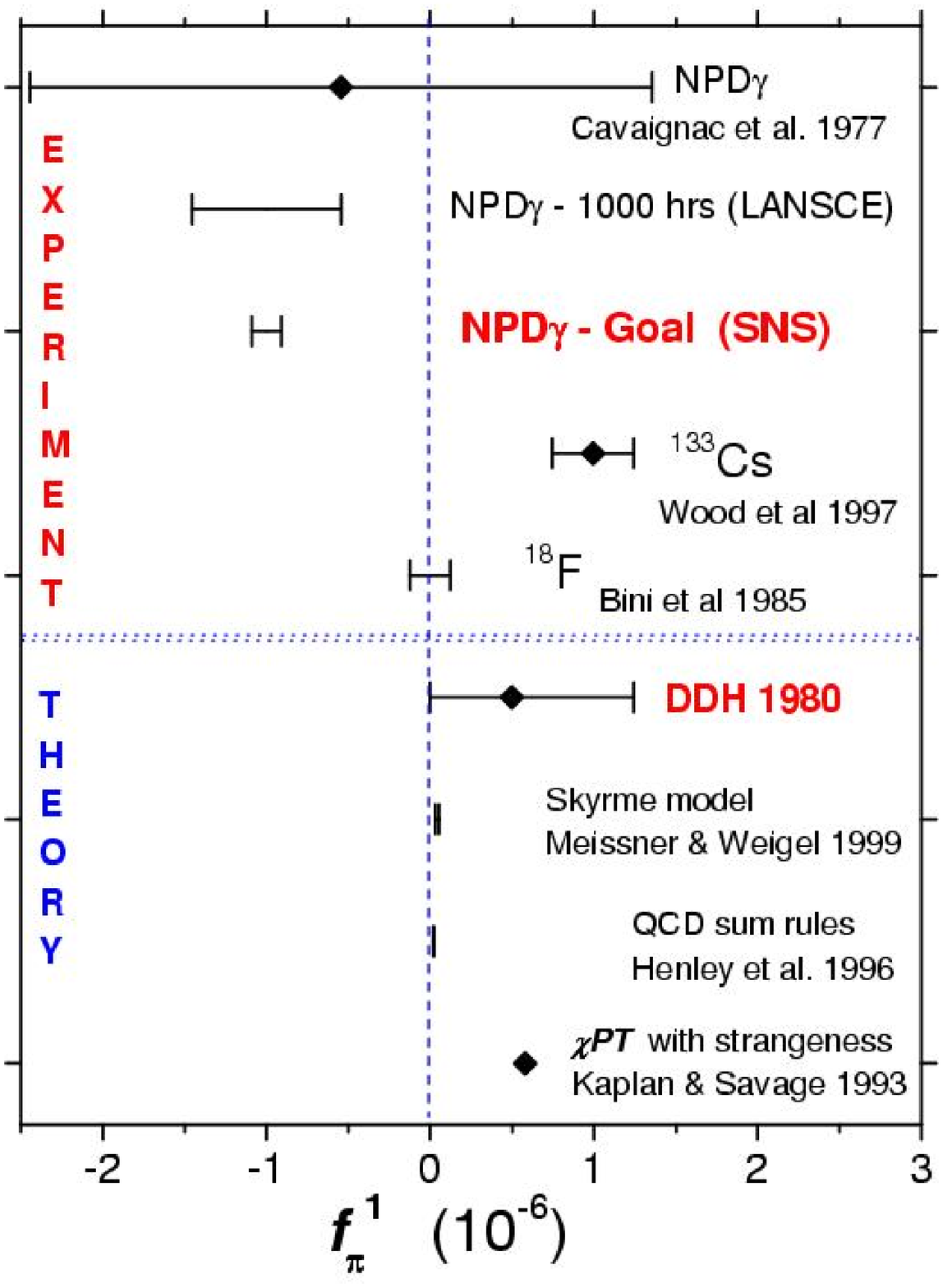}}
\vspace*{-5mm}
\caption{
a) Sketch of the polarized neutron capture process on the proton.
b) Present knowledge of the weak pion-nucleon coupling {\it f$^1_{\pi}$}.
A previous n+p$\rightarrow$d+$\gamma$ measurement gives only an upper limit.
Values extracted from the anapole moment measurement in $^{133}$Cs and the
circular photon polarization in $^{18}$F are contradictory
although nuclear effects are thought to be well understood in these systems \cite{exp}.
Precision goals for this experiment are also indicated.
Specific model calculations using effective field theory approaches yield results
within the reasonable range of the DDH approach \cite{DDH}.
}
\label{process}
\end{figure}

The high precision testing of the Standard Model in leptonic weak interactions
is in striking contrast to measurements in hadronic systems.
There, weak effects are typically 7 orders of magnitude smaller 
than the dominant strong interaction,
and observables are accessible
through parity violating (pv) phenomena.
Precise experiments that allow a
clean analysis in terms of fundamental physics parameters are missing.
Even the ``cleanest'' results are inconsistent (Fig.\ref{process}b, \cite{exp}) and 
either complicated by nuclear model dependence
or experimental difficulties.
Given the precise understanding of
electroweak physics at higher momentum,
measurements of parity violation (PV) in nucleon-nucleon ($NN$) interactions provide a tool for
testing the nucleon structure, quark-quark interactions
and chiral symmetry breaking in the non-perturbative regime.

\vspace*{-2mm}

Desplanques, Donoghue and Holstein (DDH)
developed two decades ago the 
ever-since standard
description of low-energy pv effects in $NN$ systems
by parameterization 
in terms of seven independent meson exchange couplings.
Meson exchange currents are assumed to be the appropriate description for
the low-energy $NN$ interactions, as the typical interaction length of 1.5~fm 
is much larger than the range of the heavy $Z$ and $W$ weak exchange bosons 
between pointlike leptons. 
However, uncertainties in the strong interactions permit
DDH to calculate only a broad {\it ``reasonable range''} 
for the coupling constants.
Most important, {\it f$^1_{\pi}$} describes the 
long-range part of the $NN$ potential
with a unique sensitivity to neutral currents.
Most experimentally observable processes determine a 
combination of different couplings, but
the measurement of the pv (up-down) $\gamma$-asymmetry $A_{\gamma}^{np}$ in
polarized neutron capture on the proton (Fig.\ref{process}) is unique, 
as it allows a clean determination of {\it f$^1_{\pi}$} on the few percent level
without nuclear model dependence. Only recently work has started on a systematic description of 
hadronic PV within the framework of a low-energy effective field theory,
which would finally connect the meson-exchange picture to the basic 
principles of QCD \cite{ramseymusolf}.

\begin{figure}[t]
{\includegraphics[angle= 90, height= 0.22\textheight ]{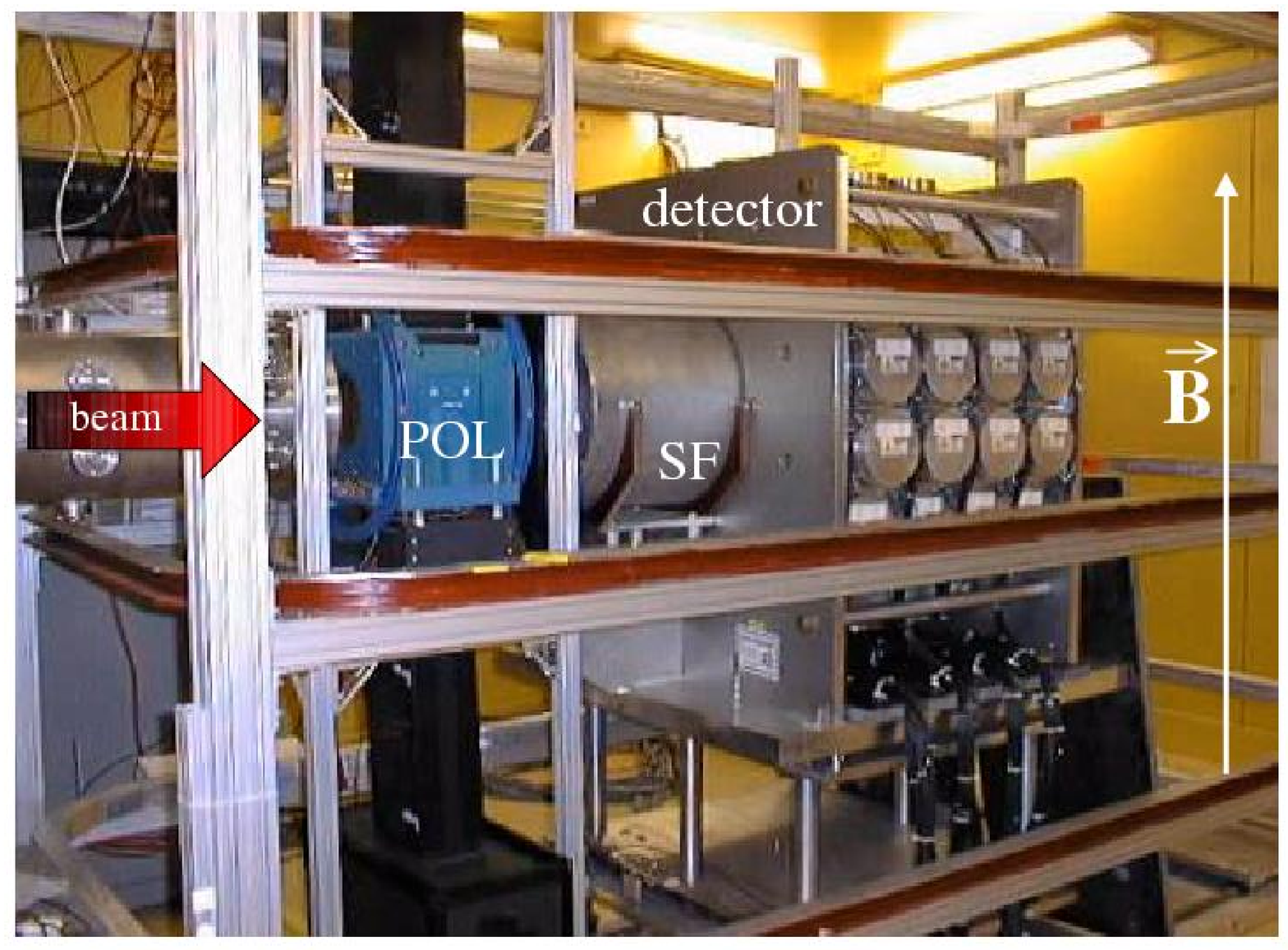}}
{\includegraphics[height= 0.22\textheight ]{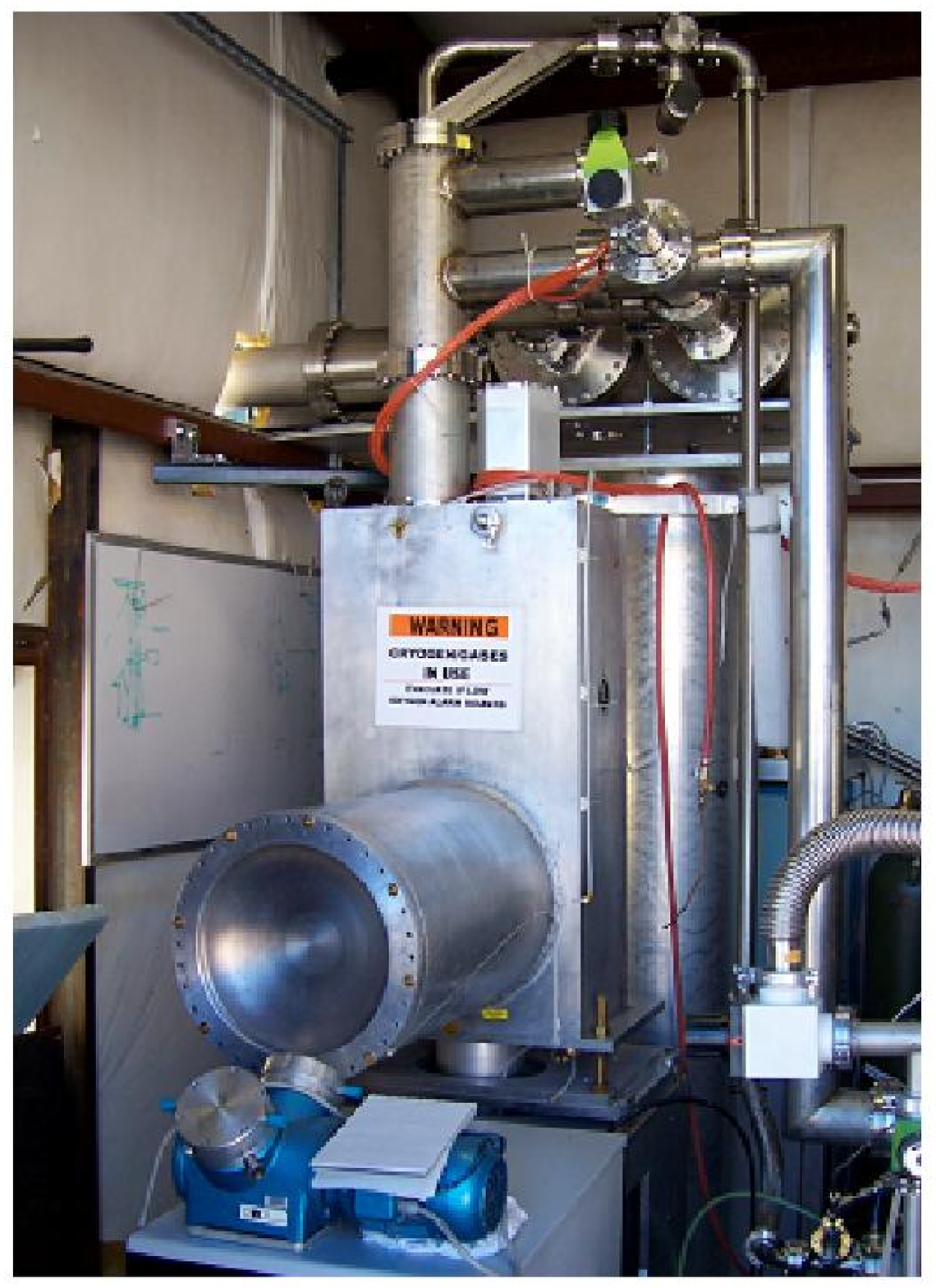}}
{\includegraphics[height= 0.23\textheight ]{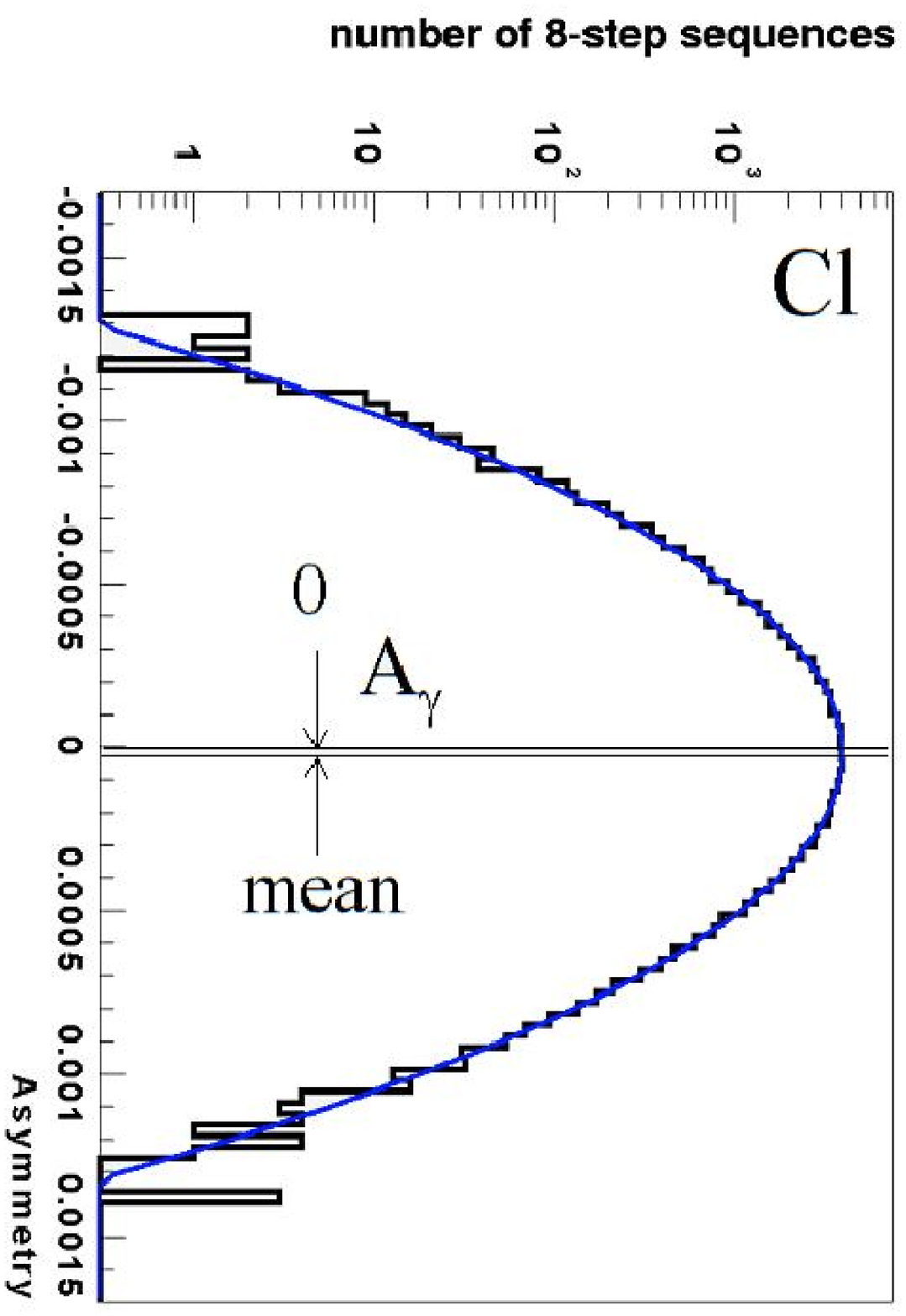}}

\vspace*{-5mm}
\caption{
The NPD$\gamma$ setup in FP12 at LANSCE: 
a) Neutron guide, $^3$He polarizer (POL), spin flipper (SF) and 
CsI detector array surrounded by the guide coils in FP12.
b) The liquid hydrogen target cryostat with the barrel-shaped target section.
c) Result for the pv asymmetry measurement in chlorine used as calibration,
as  $A_{\gamma}^{Cl}$ is 4 orders of magnitude larger than $A_{\gamma}^{np}$.
}
\label{experiment}
\end{figure}

The NPD$\gamma$ experiment sets out to measure for the first time a value
of $A_{\gamma}^{np}$ with a final precision goal of $\sim$20\% of the 
DDH ``best estimate'' value \cite{DDH}.
This requires a perfect interplay of many components
of the experimental apparatus (Fig.\ref{experiment}):

\begin{itemize}
{\small 
\item {a {\it pulsed intense cold neutron beam} delivered at LANSCE 
(and in the future at the SNS FNPB beamline), which defines
a neutron time-of-flight, and a frame definition chopper 
installed in flight path 12 (FP12);}

\item {a {\it $^3$He optically-polarized neutron spin filter} 
which selects neutrons based on the 
strong spin dependence of the cross section
for the absorption of neutrons by $^3$He;}
 

\item {a {\it resonant RF spin-flipper} to
reverse the neutron spin within a broad range of neutron energies.
Using a specific 8-step neutron spin sequence 
-- $\uparrow \downarrow \downarrow \uparrow \downarrow \uparrow \uparrow \downarrow$ --
in our measurements minimizes certain systematic effects;}

\item {a {\it 16 liter liquid para-hydrogen cryo-target}.
In the energy range up to 15 meV 
the neutron scattering cross-section in para-hydrogen 
is of the order of the neutron capture cross-section,
in ortho-hydrogen it is 2 orders of magnitude larger. 
Hence, at the time of neutron capture all spin information is 
lost. A para-hydrogen target is therefore indispensable.}

\item {an efficient {\it CsI $\gamma$-detector array} with 48 individual counters using magnetic field
insensitive vacuum photo-diodes, operating in current mode at counting statistics,
and read out by low-noise amplifiers;}

\item {a spin-polarized {\it $^3$He neutron spin analyzer} after the hydrogen target;}

\item {a {\it homogeneous 10 Gauss guide field} with field gradient $\le$ 1~mGauss/cm  
surrounding the setup to prevent Stern-Gerlach 
steering of neutrons;}

\item {{\it neutron beam monitors} to measure the beam flux and polarization;}

\item {a {\it well-shielded experimental cave} to reduce environmental noise and stray fields.}
}
\end{itemize}

The NPD$\gamma$ experiment has proven its capabilities by observation
of small pv asymmetries in setup materials
for systematic studies and by demonstrating negligible detector noise levels.
We have also performed physics studies of PV in medium heavy nuclei
(e.g. Fig.\ref{experiment}c).
NPD$\gamma$ will be ready to take hydrogen production data in early 2006.

\vspace*{-2mm}

\bibliographystyle{aipprocl} 

\end{document}